\documentclass[twocolumn,showpacs,preprintnumbers,amsmath,amssymb]{revtex4}

\usepackage{graphicx}
\usepackage{dcolumn}
\usepackage{bm}
\usepackage{amsmath}
\usepackage{amssymb}

\begin{document}

\title{Experimental Determination of $p$-wave Scattering Parameters in ultracold $^6$Li atoms}

\author{Takuya Nakasuji$^{1}$, Jun Yoshida$^{1}$, and Takashi Mukaiyama$^{1,2}$}
\affiliation{
$^{1}$\mbox{Institute for Laser Science, University of Electro-Communications, 1-5-1 Chofugaoka, Chofu, Tokyo 182-8585, Japan}\\
$^{2}$\mbox{Center for Frontier Science and Engineering, University of Electro-Communications,}\\
\mbox{1-5-1 Chofugaoka, Chofu, Tokyo 182-8585, Japan}\\
}

\date{\today}
\begin{abstract}
We report the experimental determination of the scattering parameters for a $p$-wave Feshbach resonance in a single component Fermi gas of $^6$Li atoms in the lowest spin state.
The time scale of the cross-dimensional relaxation reflects the elastic scattering rate of the atoms, and scattering parameters are determined from the scattering rate as a function of magnetic field by taking into
account the momentum distribution and inhomogeneous density profile of the atoms in a trap. Precise determination of the scattering parameters for a $p$-wave Feshbach resonance is an important step toward the realization of a $p$-wave superfluid in an ultracold atomic gas.
\end{abstract}
\maketitle

\section{\label{sec:introduction}Introduction}

Trapped fermionic atoms with tunable interactions have opened up possibilities to study novel phenomena in quantum fluids. In particular, superfluidity in  strongly-interacting fermion systems is one of the most important areas of research due to its relevance to high-$T_{\rm{c}}$ superconductivity, neutron matters, and so on. So far $s$-wave Feshbach resonances have been widely used to study the crossover between a Bose-Einstein condensate (BEC) and a Bardeen-Cooper-Schrieffer (BCS) superfluid \cite{Regal, Zwierlein}. Recently, it has become possible to derive the thermodynamic properties of  uniform fermions with a divergent $s$-wave scattering length from the thermodynamics of a harmonically trapped system, enabling the direct comparison of experimental results with many-body theories \cite{Horikoshi, Nascimbene, Ku}.

Not only can the strength and the sign (repulsive or attractive) of the interactions in an $s$-wave channel be controlled, it is possible to control the atomic interaction in a collision with non-zero angular momentum.
It is known that superfluidity with non-zero partial wave pairing arises in several condensed-matter systems \cite{Lee, Tsuei, Mackenzie}.
An ultracold atomic gas is a unique system to study $p$-wave superfluids since a fermion system with purely $p$-wave interactions can be prepared using a $p$-wave Feshbach resonance in a single component Fermi gas of atoms. Therefore, it is considered with confidence that the superfluidity emerges due to the formation of $p$-wave pairs in such a system. Furthermore, the tunability of $p$-wave interactions will enable  exploration of the evolution of superfluid properties from a BEC of $p$-wave molecules to a BCS superfluid \cite{Gurarie,Cheng,Ohashi,Iskin}. With precise control of experimental parameters and the capability of direct observation of the motions of single atoms through absorption imaging, we may be able to deepen our understanding of the microscopic mechanism of pair formation and the emergence of $p$-wave superfluidity in a strongly-interacting fermion system. 
However, strong inelastic collision losses near $p$-wave Feshbach resonances have made the realization of a $p$-wave superfluid one of the biggest challenges in the field of trapped Fermi gases \cite{Gaebler,Zhang_ENS,Chevy,Schunck,Fuchs,Inada1,Maier}.

Interactions among ultracold atoms through a $p$-wave channel can be described by the low-energy expansion of the $p$-wave scattering amplitude given by the effective-range theory \cite{Taylor}:
\begin{equation}
f_p (k)=\frac{k^2}{-\frac{1}{V(B)}+k_{\rm e} k^2 -i k^3}. \label{scattering_amplitude}
\end{equation}
Here, $V(B)$ and $k_{\rm e}$ are the scattering volume and the second coefficient in the effective-range expansion, respectively. $V(B)$ has the form of a resonance $V(B)=V_{{\rm bg}} [1+\Delta B/(B-B_{{\rm res}})]$ near a Feshbach resonance, with $V_{{\rm bg}}$, $\Delta B$ and $B_{{\rm res}}$ being the background scattering volume, resonance width, and resonance magnetic field, respectively \cite{Idziaszek}. 
Typically, the transition temperature of a $p$-wave superfluid is higher for a larger $p$-wave scattering volume \cite{You}.
Since the background $p$-wave scattering volume is not large enough to achieve a realizable critical temperature, it is important to utilize a Feshbach resonance to increase the scattering volume. 
However, a single-component Fermi gas of atoms near a $p$-wave Feshbach resonance suffers from three-body inelastic collision losses, which shows a clear contrast to the fact that a two-component Fermi gas of atoms near an $s$-wave Feshbach resonance is highly stable against three-body collision losses because of the Pauli principle.
Since the enhancement of elastic collisions and strong inelastic collision losses are inextricably linked near a Feshbach resonance, 
it is necessary to determine precisely the $p$-wave scattering parameters and find the optimum experimental conditions to realize $p$-wave superfluidity.
In this paper, we report the experimental determination of the scattering parameters for the $p$-wave Feshbach resonance in a single component Fermi gas of $^6$Li atoms in the lowest spin state at 159 G \cite{Zhang_ENS,Chevy,Schunck,Fuchs,Inada1,Maier}.
The magnetic-field dependence of the $p$-wave elastic scattering cross section is measured by a widely-used cross-dimensional relaxation method \cite{Monroe}.
By applying a theory which takes into account an inhomogeneous density profile and atomic momentum distribution, we determine the scattering parameters from the measured magnetic-field dependence of the $p$-wave elastic scattering cross sections.

This article is organized as follows. Section II describes our experimental setup for the preparation of a single-component degenerate Fermi gas of $^6$Li atoms in the lowest spin state.
In Section III, we discuss the observation of the $p$-wave Feshbach resonance from an atomic loss feature and describe how we determine the scattering parameters for the $p$-wave Feshbach resonance 
from the elastic scattering cross sections measured in the cross-dimensional relaxation method.
In Section IV, we present a summary and outlook of this work.

\section{\label{sec:introduction}Experiment}
\subsection{\label{sec:introduction}Preparation of ultracold $^6$Li atoms}

In our experiment, we employed an all-optical method of creating a degenerate Fermi gas of $^6$Li atoms in the hyperfine ground state of $|F,m_F \rangle =| 1/2, 1/2 \rangle ( \equiv |1 \rangle)$ and $|F,m_F \rangle =| 1/2, -1/2 \rangle ( \equiv |2 \rangle)$. Atoms in a magneto-optical trap were compressed by reducing the cooling laser detuning and increasing the magnetic field gradient within 20 msec, and were then transferred into a cavity-enhanced optical dipole trap \cite{Inada1,InadaBragg}.
A cavity-enhanced optical trap was realized by placing a 400 mm cavity in a confocal configuration. The 1064 nm laser power in the cavity was enhanced by factor of 100 to create a trap depth of $k_B \times$ 1 mK with a beam waist of 260 $\mu$m. We then transferred the atoms into a single-beam optical trap with a beam waist of 35 $\mu$m. A radio-frequency (RF) field was applied to make the population equal in the $|1 \rangle$ and $|2 \rangle$ states. Evaporative cooling was performed at 300 G where the absolute value of the scattering length between $|1 \rangle$ and $|2 \rangle$ states is at a local maximum. A temperature of $T/T_F \sim 0.1$ with $6 \times 10^5$ atoms was achieved.

To prepare a single-component Fermi gas of $^6$Li atoms in the $ |1 \rangle$ state, we adiabatically ramp the magnetic field to 215 G where the $ |2 \rangle - |2 \rangle$ $p$-wave Feshbach resonance is located, and remove the atoms in the $ |2 \rangle $ state by enhanced three-body collision loss for $ |2 \rangle $ state atoms. We checked that we prepared atoms purely in the state $|1\rangle$ using a conventional Stern-Gerlach measurement. The number of remaining atoms in $|2\rangle$ was confirmed to be less than 0.1\% of the number of atoms in $|1\rangle$ state .
After the preparation of pure $ |1 \rangle$ state atoms, we set the magnetic field to $ |1 \rangle - |1 \rangle$ $p$-wave Feshbach resonance at around 159 G. 

Since the current provided by a commercial power supply typically has ripple noise in the order of 0.1 \%, the magnetic field created by the current has the same amount of fluctuation. In order to reduce the magnetic field fluctuation, we stabilize the current running in the coils by controlling the current in bypass circuit added in parallel to the coils. With the stabilization, the current fluctuation is reduced to $5 \times 10^{-5}$ of the current value, which corresponds to a magnetic field fluctuation of 8 mG.

\begin{figure}[b]
\includegraphics[scale = 0.58]{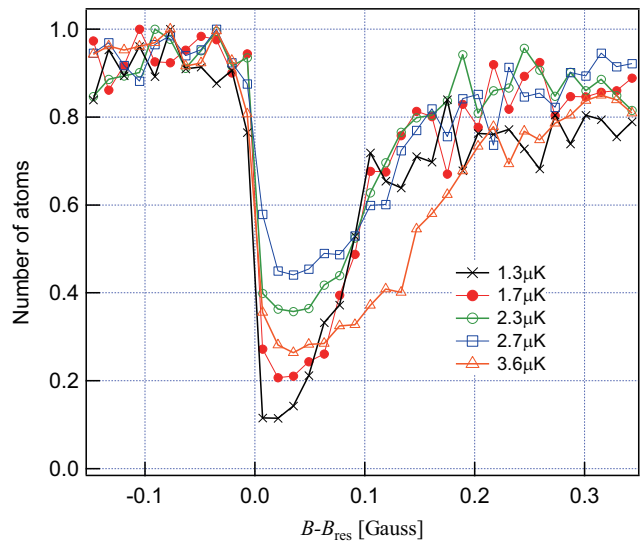}\\
\caption{Fraction of atoms remaining after 50 ms holding time near the $|1 \rangle - |1 \rangle $ $p$-wave Feshbach
resonance as a function of magnetic field strength. The atomic loss feature becomes broader at the higher magnetic field side at higher temperature 
because of the wider spread of the momentum distribution.
The sharp edge at the lower magnetic field side of the atomic loss feature indicates the position of the Feshbach resonance.\label{atomic_loss}}
\end{figure}

\subsection{Determination of the $p$-wave Feshbach resonance from atomic loss features}

Figure \ref{atomic_loss} shows the atomic loss measured when atoms in state $|1\rangle$ are held in a trap for various magnetic fields. Since atoms are purely in the lowest energy state $|1\rangle$, atomic loss is only due to three-body collisions. The vertical axis shows the fraction of atoms remaining after the 50 ms holding time, and the horizontal axis shows the magnetic field detuning from the $p$-wave Feshbach resonance at $B_{\rm res}= 159.17(5)$ G \cite{Zhang_ENS,Chevy,Schunck,Fuchs,Inada1,Maier}. $B_{\rm res}$ was determined from the frequency to drive an RF transition from $|1\rangle$ to $|2\rangle$ at the Feshbach resonance magnetic field. Data for five different temperatures: 1.3 $\mu$K, 1.7 $\mu$K, 2.3 $\mu$K, 2.7 $\mu$K and 3.6 $\mu$K, are shown with black crosses, red closed circles, green open circles, blue open squares and orange open triangles, respectively. Since atom pairs with high relative momentum (namely high collision energy) have a higher Feshbach resonance field, an asymmetry appears in the loss feature due to the spread of atomic energy distribution. When we increase the temperature of the atoms, the atomic loss feature broadens only toward the higher magnetic-field side. The edge of the loss feature in the lower magnetic-field side is always sharp for all temperatures as shown in Fig. \ref{atomic_loss}.
The sharp edge of the loss feature corresponds to the $p$-wave Feshbach resonance for the atomic pairs with zero kinetic energy.
We determine the $p$-wave Feshbach resonance magnetic field from the sharp edge at the lower magnetic-field side of the atomic loss feature.

\subsection{Measurement of elastic scattering cross section}

In order to determine the unknown scattering parameters, $V_{\rm bg}, \Delta B$, and $k_{\rm e}$ in the scattering amplitude function (Eq.~(\ref{scattering_amplitude})), we measure the $p$-wave elastic scattering cross section near the $|1 \rangle - |1 \rangle $ $p$-wave Feshbach resonance. The elastic scattering cross section is given by the scattering amplitude as $\sigma = 12 \pi |f_p (k)|^2$. In our experiment, we determine the elastic scattering cross section using a cross-dimensional relaxation method \cite{Monroe}. In this measurement, we first prepare a dipole-trap laser with an elliptical beam waist so that the trap frequencies in the two radial directions are different by 20 \%. When we modulate the trapping-laser intensity with a frequency exactly twice the trap frequency in one of two radial directions, we can excite the atomic motion in the resonant radial direction, such that there is additional kinetic energy in one direction, resulting in an anisotropic momentum distribution. At this time, absorption images after the time-of-flight (TOF) show anisotropic expansion of the cloud. When we hold the atoms in the trap after the parametric excitation, we see the time evolution of the aspect ratio of the cloud reaching unity. This is because the inter-atomic scattering transfers the kinetic energy from one direction to another, forcing the system to reach thermal equilibrium. The time of this thermalization is determined by the elastic collision cross section, and therefore the measurement of this energy-transfer rate as a function of the magnetic field gives us information about the scattering parameters.
It is known that a p-wave Feshbach resonance has a doublet structure which  originates from a magnetic dipole-dipole interaction between two atoms in a p-wave molecular state \cite{Ticknor}. Since the splitting of the two resonances is predicted to be small (in the order of mG \cite{Zhang_ENS}) compared with the width of the resonance and smearing of the resonance in the magnetic field due to the momentum spread of the atoms, we treat the p-wave Feshbach resonance as a single resonance in this work.

\begin{figure}[t]
\includegraphics[scale = 0.85]{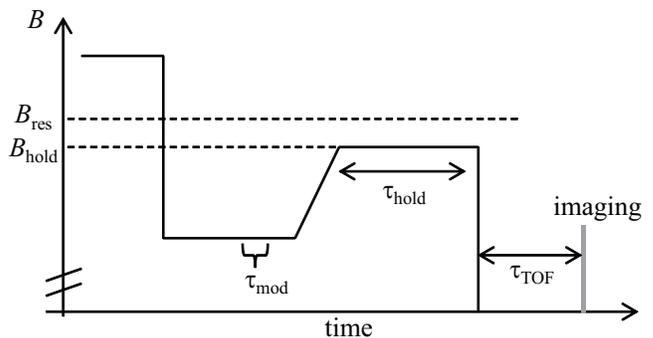}\\
\caption{Experimental time chart of the magnetic field. The magnetic field is swept quickly toward the lower field side of the resonance ($B_{{\rm res}}$) to avoid adiabatic creation of $p$-wave molecules. Then, the trapping laser intensity is modulated for 3~msec with twice the trap frequency in one direction to add kinetic energy. The magnetic field is set close to the resonance ($B_{{\rm hold}}$) where the elastic collision rate is measured. After holding the atoms at $B_{{\rm hold}}$, the magnetic field is quickly swept to zero when atoms are released from the trap.\label{time_chart}}
\end{figure}

Figure \ref{time_chart} shows the experimental sequence for the magnetic field in the measurement of the elastic scattering cross sections. After the preparation of the atoms in the $|1 \rangle $ state, the magnetic field is above the $p$-wave Feshbach resonance $B_{{\rm res}}$. We sweep the magnetic field to somewhere far below $B_{{\rm res}}$ (where the inter-atomic $p$-wave interaction is negligibly small) as fast as possible to avoid adiabatic creation of $p$-wave Feshbach molecules \cite{Inada1}. Then the trapping-laser intensity is modulated with twice the trap frequency in one direction for $\tau_{\rm mod}$ = 3~msec which increases the kinetic energy of the atoms in one of the two radial directions. After holding the system for 5~msec to let the atomic motion dephase, the magnetic field is changed quickly to $B_{{\rm hold}}$ where the thermalization time is measured. Measurement of the atomic momentum distribution is done by  absorption imaging after releasing the atoms from the trap when the magnetic field is turned off.

\begin{figure}[b]
\includegraphics[scale = 0.85]{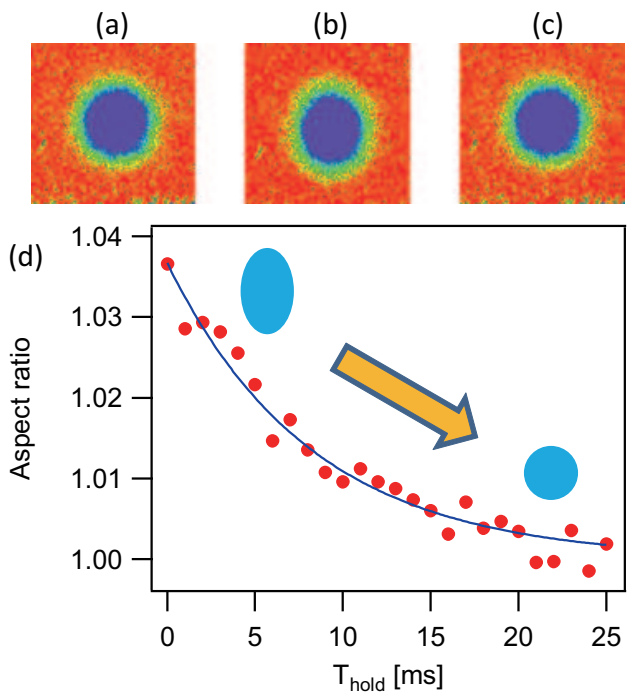}\\
\caption{Time evolution of the anisotropy of the momentum distribution measured after a 4 msec TOF. (a) Momentum distribution measured before the parametric excitation. Atoms are seen to expand isotropically. (b) Momentum distribution measured just after the parametric excitation. The momentum distribution is wider in the vertical direction than in the horizontal direction in the image. (c) Momentum distribution after reaching thermal equilibrium due to inter-atomic scattering. Momentum distribution is observed to be isotropic again. (d) A typical time evolution of an aspect ratio as a function of hold time $T_{\rm hold}$. \label{aspect_ratio}}
\end{figure}

Figures \ref{aspect_ratio}(a), \ref{aspect_ratio}(b), and \ref{aspect_ratio}(c) show the momentum distribution observed in TOF images (a) before applying the parametric excitation, (b) just after applying the excitation, and (c) after holding the atoms for longer than the thermalization time, respectively. As shown in Fig. \ref{aspect_ratio}(b), the expanded cloud becomes elliptical, indicating that the resonant modulation of the trapping laser intensity increases the kinetic energy of the atoms in the vertical direction. After holding the atoms long enough to reach thermal equilibrium due to inter-atomic $p$-wave scattering, the kinetic energies in the vertical and horizontal directions become equal and the momentum distribution of the atoms becomes isotropic again (Fig. \ref{aspect_ratio}(c)). Figure \ref{aspect_ratio}(d) shows the typical time evolution of the aspect ratio of the atomic cloud in the absorption images. The red markers show the experimental data and the solid curve shows the fitting result with an exponential decay curve to the data. The aspect ratio gradually decreases due to thermalization by $p$-wave scatterings and finally reaches unity after a long holding time. The time evolution of the aspect ratio can be described by $1+A \exp (-\gamma T_{\rm hold})$, where $A$ is an amplitude of excitation and $\gamma$ is a thermalization rate. The ratio of kinetic energies in the two orthogonal directions is proportional to the square of the aspect ratio, $E_{\rm V} / E_{\rm H}  \simeq 1+2A \exp (-\gamma T_{\rm hold})$. Therefore the thermalization rate $\gamma$ actually indicates the rate of energy transfer from one radial direction to the other.

\begin{figure}[t]
\includegraphics[scale=0.7]{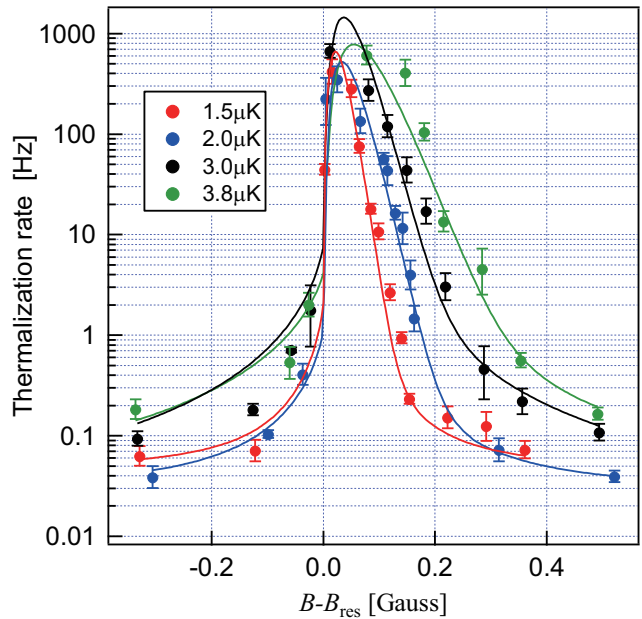}
\caption{Thermalization rate as a function of magnetic field measured near the $p$-wave Feshbach resonance for four different temperatures. 
Markers show the experimental data and solid curves show the fitting result using Eq.~(\ref{fitting_function}). 
The zero of the magnetic field detuning is determined from the loss measurement shown in Fig. \ref{atomic_loss}.\label{energy_transfer}}
\end{figure}

Figure \ref{energy_transfer} shows the measured thermalization rate as a function of the magnetic field detuning. The measurements were done for the atomic clouds at four different temperatures. Red, blue, black and green markers show the results using the atoms at 1.5~$\mu K$, 2.0~$\mu$K, 3.0~$\mu$K, and 3.8~$\mu$K, respectively. The zero of the magnetic-field detuning $B_{\rm res}$ was determined from the atomic loss measurement shown in Fig. \ref{atomic_loss}. The thermalization rate drastically increased by more than three orders of magnitude above zero magnetic-field detuning, and the range of the magnetic field where fast thermalization is observed gets wider at higher temperature. Error bars include both the statistical errors and the effect of underestimation of the rate due to atomic losses within the thermalization timescale \cite{loss}. 
In this experiment, $T/T_{\rm F}$ is always chosen to be larger than 1.0 so that the momentum distributions of the atoms are well described by Boltzmann distributions.

In order to determine the parameters in the scattering amplitude from the data shown in Fig. \ref{energy_transfer}, we fit the data to a theoretical curve which takes into account the thermal averaging effect \cite{Ticknor,Burke}. The number of two-body scattering events per unit volume per unit time is given by $n \times n \sigma (E) v_r$, where $n$, $\sigma (E)$ and $v_r$ are the atomic density, the elastic scattering cross section for the relative scattering energy $E$  and the relative velocity, respectively. The collision energy $E$ is defined to be $E=\frac{1}{2}\mu v_r^2$ with $\mu$ being a reduced mass. The elastic scattering cross section is expressed in terms of the scattering amplitude as $\sigma (E) = 12\pi |f_p (E)|^2$. 
The experimental data shown in Fig. \ref{energy_transfer} is the rate of energy transfer from one radial direction to the other, which can be calculated by the product of the collision energy $E$ and the number of two-body scattering events per unit time $ n^2 \sigma (E) v_r $. The energy transfer rate is given by $\gamma = \frac{2}{\alpha} \langle n^2 \sigma (E) v_r E \rangle /k_{\rm B}T$ \cite{Burke}, where  $\alpha =4.1$ is the average number of $p$-wave scatterings per atom required for thermalization. 
The bracket indicates the thermal averaging of the value, which is obtained by integrating over phase space after multiplying by the phase space distribution function \cite{Burke}. 
Since $n$ is the function of the position $\vec{x}$, and $\sigma (E)$, $v_r = p_r/\mu$, and $E= p_r^2/2\mu$ are the functions of the relative momentum $p_r$, the value of $\langle n^2 \sigma (E) v_r E \rangle $ can be written as \cite{Burke}
\begin{eqnarray}
\langle n^2 \sigma (E) v_r E \rangle _{\vec{x}, \vec{p}} & = & \langle n^2 \rangle _{\vec{x}} \langle \sigma (E) v_r E \rangle _{\vec{p}} \nonumber \\
& = & n_{{\rm mean}} \times \sqrt{\frac{8}{\mu \pi}} \left( k_{\rm B} T \right)^{-3/2} 12\pi \nonumber \\ 
& & \times \int |f_p (E)|^2 E^2 \exp (-\beta E) dE, \label{fitting_function}
\end{eqnarray}
where $n_{{\rm mean}}=\frac{1}{N}\int n(\vec{x})^2 d\vec{x}$ is the mean density. 
The mean density is determined from the atomic density profile.

The red, blue, black, and green curves in Fig. \ref{energy_transfer} are the fitting results for the experimental data with the atoms at $T=1.5 \mu K$, 2.0~$\mu$K, 3.0~$\mu$K, and 3.8~$\mu$K, respectively.
When we fit the data with the theoretical curve, $V_{\rm bg}\Delta B$, $k_{\rm e}$ and the offset value of the thermalization rate $\gamma_{\rm offset}$ are taken as the fitting parameters. $V_{\rm bg}$ and $\Delta B$ are only determined as a product of two values because the magnetic field measurement region is small compared with $\Delta B$ and so the first term in the parentheses (namely the non-resonant part of the scattering volume) in $V(B)=V_{{\rm bg}} [1+\Delta B/(B-B_{{\rm res}})]$ is negligible \cite{footnote1}. The offset values for the thermalization rate are finite because the anharmonicity of the trapping potential causes mixing of the motions in the two radial directions which in turn creates an offset thermalization rate.
This anharmonicity effect is larger when the trapping potential is tighter. Since we control the temperature of the atoms by changing both the tightness of the trap potential and the extent to which we conduct evaporative cooling, 
the effect of anharmonicity depends on the experimental conditions as does the offset value of the thermalization rate.
From the fitting, we find $V_{\rm bg}\Delta B=-2.8 (3) \times 10^6 a_0^3$ [G] and $k_{\rm e}=0.058 (5) a_0^{-1}$. These values agree reasonably well with the theoretical calculation results of $V_{\rm bg}\Delta B=-1.45\times10^6 a_0^3$ [G] \cite{Julienne}, $-1.7\times10^6 a_0^3$ [G] \cite{Austen} and $k_{\rm e}=0.048 (3) a_0^{-1}$ \cite{Julienne}. 
If we use the predicted value of $\Delta B = 40$ [G] \cite{Austen}, we get $a_{\rm bg}=V_{\rm bg}^{1/3}= -41 a_0$. This value agrees reasonably well with the theoretically predicted values of $-35 a_0$ \cite{You}, $-38 a_0$ \cite{Gautam}, and $-35.3 a_0$ \cite{Zhang}. The measured magnetic field dependence of the scattering volume is also consistent with the calculation by Lysebo et al. \cite{Lysebo}.

\section{conclusion and outlook}

We have experimentally determined the scattering parameters for a $p$-wave Feshbach resonance in a single component Fermi gas of $^6$Li in the lowest spin state.
The cross-dimensional relaxation technique was used to measure the thermalization time of the excitation applied to a cloud. From the analysis which takes into account an inhomogeneous density profile and momentum distribution of the atoms,
we were able to derive the unknown scattering parameters used to form an expression for the elastic collision cross sections. The determined parameters show reasonable agreement with theoretical calculations.
Since strong atomic loss near a $p$-wave Feshbach resonance prevents us from realizing a $p$-wave superfluid in an atomic gas, it is important to know the magnetic-field dependence of the scattering volume and inelastic loss rate precisely.

Since the magnetic field dependence of the scattering volume has been determined, we may be able to experimentally confirm the universal property in three-body collisions. Suno et al. predicted that the three-body collision coefficient has a $|V(B)|^{8/3}$ dependence in an identical fermion system, in contrast to the $|a|^4$ dependence in an identical boson system \cite{Suno}. Recently, Nishida et al. predicted a ``super Efimov effect'' which is expected to appear as an enhancement of three-body collision losses in a two-dimensional system \cite{Nishida}. 
The scattering parameters determined in this work will help us to experimentally confirm such unique features in three-body collisions.

Finally, a p-wave Feshbach resonance is composed of two resonances split by the magnetic dipole-dipole interaction. In this work, we analyzed the data based on the theoretical prediction that the splitting is much smaller than the features of the enhancement of the cross-dimensional relaxation. To identify such a fine doublet structure of the $p$-wave Feshbach resonance, an optical lattice can be used to confine the atoms in one- or two-dimensions and realize the situation where only one of the two resonances become effective.
There is also a theoretical proposal that trapping atoms in two dimensions helps to achieve a higher good-to-bad collision ratio near a $p$-wave Feshbach resonance \cite{Levinsen} and is a promising means of realizing $p$-wave superfluids in ultracold atomic gases. This will be a future challenge.

\section{acknowledgments}

We would like to thank T. Kaneda for his experimental assistance. We also would like to thank P. Julienne for providing us with the 
calculation results and for stimulating discussions.
This work is supported in part by the Grants-in-Aid for Scientific Research on Innovative Areas from MEXT (No.2404: 24105006)


\begin{thebibliography}{99}
\bibitem{loss} Atomic loss during the thermalization is confirmed to be less than 30 \% of the initial number of atoms. The error bars shown in Fig. \ref{energy_transfer} include the effect of underestimation of the rate 
arising from the atomic loss.
\bibitem{footnote1} It was not obvious that the non-resonant part of the scattering volume becomes negligibly small in the magnetic field range in this work. 
However, we could not reproduce the experimental data if we choose a small $\Delta B$ with which off-resonant part has non-negligible contribution to the scattering volume.
\end{thebibliography}
\end{document}